\documentclass{PoS}

\usepackage{amsmath,amssymb,bbm}

\newcommand{\krknlo}{{\textsf{KrkNLO}}}
\newcommand{\mcatnlo}[1]{\textsf{MC@NLO#1}}
\newcommand{\powheg}[1]{\textsf{POWHEG#1}}
\newcommand{\sherpa}[1]{\textsf{Sherpa#1}}
\newcommand{\mcfm}[1]{\textsf{MCFM#1}}
\newcommand{\herwig}[1]{\textsf{Herwig #1}}
\newcommand{\herwigver}[1]{\textsf{Herwig++~2.7.0#1}}
\newcommand{\sherpaver}[1]{\textsf{Sherpa~2.0.0#1}}
\newcommand{\herwigsiedem}[1]{\textsf{Herwig~7.0 #1}}

\newcommand{\mc}{\textrm{MC}}
\newcommand{\lo}{\textrm{LO}}
\newcommand{\nlo}{\textrm{NLO}}

\newcommand{\qbar}{{\bar q}}
\newcommand{\F}{{_{F}}}
\newcommand{\B}{{_{B}}}
\newcommand{\msbar}{\overline{MS}}

\title{  New simpler method of matching NLO corrections 
           with parton shower Monte Carlo\thanks{This work is partly supported by 
  the Polish National Science Centre grant UMO-2012/04/M/ST2/00240.}}

\ShortTitle{KrkNLO method}

\author{\speaker{S. Jadach}\\
  Institute of Nuclear Physics, Polish Academy of Sciences,\\
  ul.\ Radzikowskiego 152, 31-342 Krak\'ow, Poland\\
 E-mail: \email{stanislaw@jadach@ifj.edu.pl}}

\author{ W.\ P\l{}aczek\\
  Marian Smoluchowski Institute of Physics, Jagiellonian University,\\
  ul.\ {\L}ojasiewicza 11, 30-348 Krak\'ow, Poland}

\author{  S.\ Sapeta\thanks{On leave of absence from INP PAS, Krak\'ow, Poland.} \\
  Theoretical Physics Department, CERN, Geneva, Switzerland}

\author{  A.\ Si\'odmok\thanks{On leave of absence from INP PAS, Krak\'ow, Poland. }\\
  Theoretical Physics Department, CERN, Geneva, Switzerland}

\author{ M.\ Skrzypek\\
  Institute of Nuclear Physics, Polish Academy of Sciences,\\
  ul.\ Radzikowskiego 152, 31-342 Krak\'ow, Poland}

\abstract{Next steps in development of the \krknlo\
 method of implementing NLO QCD corrections to hard processes
 in parton shower Monte Carlo programs are presented.
 This new method is a simpler alternative to other well-known
 approaches, such as \mcatnlo\ and \powheg.
 The \krknlo\ method
 owns its simplicity to the use of parton distribution
 functions (PDFs) in a new,  so-called Monte Carlo (MC), factorization scheme
 which was recently fully defined for the first time.
 Preliminary numerical results for the Higgs-boson production process
 are also presented.
}

\FullConference{Loops and Legs in Quantum Field Theory\\
		24--29 April 2016\\
		Leipzig, Germany\\
		\vspace{5mm}
		{\bf IFJPAN-IV-2016-15}
		\vspace{-5mm}\\
}
\begin{document}

\section{Introduction}

The first idea of the of the \krknlo{} methodology
was suggested at Ustro\'n 2011 Conference~\cite{Skrzypek:2011zw}
as a byproduct of another study on introduction
of NLO corrections in the QCD evolution of parton distributions
using Monte Carlo (MC) methods.
This proposal  was originally formulated for the process of $Z/\gamma$ production
in proton--proton collisions, i.e.\ the Drell--Yan (DY) process, 
and was limited to multi-gluon emission out of quarks (gluonstrahlung).
However, numerical implementation was missing.
The first numerical validation of the above \krknlo{} method
was presented shortly in ref.~\cite{Jadach:2012vs}
within the so-called ``Double-CMC'' toy-model 
parton shower (PS) Monte Carlo (MC),
still for gluonstrahlung only.

A more complete theoretical discussion of the \krknlo{} scheme,
with an explicit introduction of the parton distribution functions (PDFs)
in the MC factorization scheme, for DY and electron--proton deep-inelastic scattering (DIS)
was provided in ref.~\cite{Jadach:2011cr},
but the MC implementation was still done 
on top of the not-so-realistic Double-CMC PS,
and for gluonstrahlung in DY process only.

The first realistic implementation of the \krknlo{} method
was presented in ref.~\cite{Jadach:2015mza}.
It was done on top of the \sherpa~\cite{Gleisberg:2008ta} and \herwig~\cite{Bahr:2008pv,Bellm:2013lba,Gieseke:2012ft} PS MCs,
implementing the complete NLO QCD corrections for the DY process, 
including gluon--quark transitions.
In this work the \krknlo{} numerical results were compared with
these from
\mcatnlo~\cite{Frixione:2002ik} and 
\powheg~\cite{Nason:2004rx} implementations.
Also comparisons with the NLO fixed-order calculations of
\mcfm~\cite{MCFM} and
with the NNLO ones of ref.~\cite{Catani:2009sm}
were presented.

In the following we shall elaborate on:
(i) a mature definition of PDFs in the MC factorization scheme,
(ii) universality of the MC factorization scheme,
(iii) application to the Higgs-production process.
These issues are also covered in the
recent ref.~\cite{Jadach:2016acv}.

\section{\krknlo\ method}

\begin{figure}[h]
  \centering
  \includegraphics[height=0.25\textwidth]{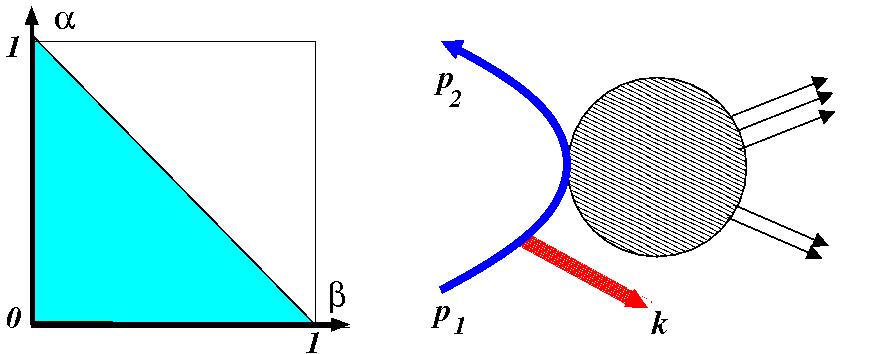}
  \caption{\sf
   Kinematics of the $g + g \longrightarrow H, H + g$ process.
   }
  \label{fig:SudakDY}
\end{figure}

A recipe of the \krknlo\ method of
introducing NLO QCD corrections
for spin-zero Higgs-boson production in proton--proton collisions
is unbelievably simple:
take an event generated by a LO-type parton shower MC as it is
and apply a simple positive well-behaved weight
\begin{itemize}
\item[a)]
for the subprocess
 $g + g \longrightarrow H$:
\begin{equation}
 W^{\mc}_{VS}(\alpha,\beta)
  = 1+\Delta_{VS}^{\mc},
\label{eq:WVS}
\end{equation}
where
\begin{equation}
 \Delta_{VS}^{\mc}
= \frac{\alpha_s}{2\pi}\, 2C_A \left[ \frac{473}{72} + \frac{2\pi^2}{3} 
 - \frac{T_f}{C_A}\frac{59}{36}\right],
\label{eq:DeltaVS}
\end{equation}
\item[b)]
for the subprocess
 $g + g \longrightarrow H + g$:
\begin{equation}
 W^{\mc}_{gg}(\alpha,\beta)
  =\frac{1+z^4 + \alpha^4 + \beta^4}{1 + z^4 + (1 - z)^4}
    \quad \leq 1,
\label{eq:Wgg}
\end{equation}
\item[c)]
for the subprocess $g + q \longrightarrow H + q$:
\begin{equation} 
   W^{\mc}_{gq}(\alpha,\beta) =
   \frac{1 + \beta^2}{1 + (1-z)^2}  \quad \leq 1.
\label{eq:Wgq}
\end{equation}
\end{itemize}
In the above we use Sudakov variables defined as follows:
\begin{equation}
  \alpha=\frac{kp_2}{p_1 p_2},\quad
  \beta =\frac{kp_1}{p_1 p_2},\quad
  \alpha + \beta = 1 - z  \leq 1,\quad
  \alpha\geq 0,\; \beta\geq 0.
\label{eq:SudVar}
\end{equation}
They are visualized in Fig.~\ref{fig:SudakDY}.
The immediate question is:  which $\alpha$, $\beta$ -- for which parton?
The answer is simple: these of the parton with maximum $k_T$,
i.e.\ for PS MC with the $k_T$-ordering, the ones of 
the 1st parton in the backward evolution.
For the angular ordering, a parton with maximum $k_T$ 
needs to be found in the MC event.

The above NLO correcting weight is particularly simple 
because it is independent of the momenta of the Higgs-decay products,
due to spin-zero of the Higgs boson.
However, the NLO weight is also quite simple in the following case of
gluonstrahlung in the $Z/\gamma$ production process:
\begin{eqnarray}
 d\sigma^{\nlo}_{n}
&=&
 \Big(1+\Delta_{VS} 
       +\sum_{i=1}^{n} W^{[1]}_{q\qbar}(\alpha_i,\beta_i)
 \Big)
  d\sigma^{\lo}_{n},
\\
  W^{[1]}_{q\qbar} 
&=&
   \frac{d^5\bar\beta_{q\qbar}}{d^5\sigma^{\lo}_{q\qbar}}
   = \frac{d^5\sigma^{\nlo}_{q\qbar} -d^5\sigma^{\lo}_{q\qbar} }%
          {d^5\sigma^{\lo}_{q\qbar}},
~~~~~~
\Delta_{VS}^{q\bar q} =  
      \frac{\alpha_s}{2\pi} C_F \left[\frac{4}{3}\pi^2 -\frac{5}{2}\right],
\\
d^5\sigma^{NLO}_{q\qbar}(\alpha,\beta, \Omega)
&=&
\frac{C_F \alpha_s}{\pi}\;
\frac{d\alpha d\beta}{\alpha\beta}\;
\frac{d\varphi}{2\pi}\;
d\Omega
\Bigg[
 \frac{d\sigma_0(\hat{s},\theta_\F)}{d\Omega}
 \frac{(1-\beta)^2}{2}
+\frac{d\sigma_0(\hat{s},\theta_\B)}{d\Omega}
 \frac{(1-\alpha)^2}{2}
\Bigg],
\\
d^5\sigma^{\lo}_{q\qbar}(\alpha,\beta, \Omega)
&=&
d^5\sigma^\F_{q\qbar} + d^5\sigma^\B_{q\qbar}
=\frac{C_F \alpha_s}{\pi}\;
 \frac{d\alpha d\beta}{\alpha\beta}\;
 \frac{d\varphi}{2\pi} d\Omega\;
 \frac{1+(1-\alpha-\beta)^2}{2}
 \frac{d\sigma_{0}}{d\Omega}\big(\hat{s},\hat\theta \big),
\end{eqnarray}
The above formula is well-suited for PS MC with
the angular ordering.
Here, the MC weight sums up ``democratically''
contributions from all gluons, picking up automatically
the contribution from the gluon with the maximum transverse momentum $k_T$.
Alternatively, we may find out which gluon has maximum $k_T$
and reduce $\sum_{i=1}^{n}$ just to one term.
In this way the \krknlo\ scheme exploits the Sudakov suppression as in \powheg,
but there is no need of truncated or vetoed showers
necessary in \powheg\  for angular ordering.

In the case of averaging over the angles of  $Z/\gamma$ decay products,
the NLO weight is even simpler.
For the subprocess $q + \bar{q} \longrightarrow Z$ it reads
\begin{equation}
W^{\mc}_{VS}(\alpha,\beta) = 
  1+\Delta_{VS}^{\mc},\qquad
 \Delta_{VS}^{\mc}=  \frac{\alpha_s}{2\pi}\, 
 \left(\frac{4\pi^2}{3} + \frac{1}{2}\right),
\label{eq:WDYVS}
\end{equation}
for $q + \bar{q} \longrightarrow Z + g$ one gets
\begin{equation}
W^{\mc}_{q\bar{q}}(\alpha,\beta) = 
\left\langle 
\frac{|{\cal M}_{q\bar{q}\rightarrow Zg}^{\nlo}|^2}%
     {|{\cal M}_{q\bar{q}\rightarrow Zg}^{\mc}|^2} 
\right\rangle_{\rm Z~decay}\!\!\!\!\!
= 1 - \frac{2\alpha\beta}{1+ z^2} \quad \leq 1,
\label{eq:WDYqq}
\end{equation}
and for $q + g \longrightarrow Z + q$  one employs
\begin{equation}
W^{\mc}_{qg}(\alpha,\beta) = 
\left\langle 
\frac{|{\cal M}_{qg\rightarrow Zq}^{\nlo}|^2}{|{\cal M}_{qg \rightarrow Zq}^{\mc}|^2} 
\right\rangle_{\rm Z~deday}
=  1 + \frac{\alpha(\alpha + 2z)}{z^2 + (1 - z)^2} \quad < 3.
\label{eq:WDYqg}
\end{equation}

Obviously, the \krknlo\ recipe is dramatically simpler 
than the \powheg\ or \mcatnlo\ methods.
However, several non-trivial conditions have to be met
before it can be applied. The most important requirements are:
\begin{itemize}
\item
PDFs in the MC factorization scheme -- the absolute must!
\item
LO PS MC must 
cover the entire NLO phase space.
\item
LO PS MC has to reproduce precisely all soft and collinear
singularities of the NLO level --
a non-trivial requirement for LO processes with $\geq 3$ colored legs.
\end{itemize}

\section{Monte Carlo factorization scheme}

The LO PDFs in the MC factorization scheme in terms of the PDFs in the $\msbar$ factorization scheme
are defined by
\begin{equation}
  \begin{bmatrix}  
     q(x,Q^2) \\ \bar{q}(x,Q^2) \\ G(x,Q^2) 
  \end{bmatrix}_{\mc}
=
 \begin{bmatrix}  
     q \\ \bar{q} \\ G 
 \end{bmatrix}_{\msbar}
+
\int\! dz dy
 \begin{bmatrix}
    K^{\mc}_{qq}(z)     & 0                       &  K^{\mc}_{qG}(z)  \\ 
    0                 & K^{\mc}_{\bar{q}\bar{q}}(z)  &  K^{\mc}_{\bar{q}G}(z) \\ 
    K^{\mc}_{Gq}(z)     & K^{\mc}_{G\bar{q}}(z)       &  K^{\mc}_{GG}(z)
 \end{bmatrix}
 \begin{bmatrix}  
     q(y,Q^2) \\ \bar{q}(y,Q^2) \\ G(y,Q^2) 
 \end{bmatrix}_{\msbar}
  \!\!\!\!   \delta(x-yz),
\label{eq:msbar2mc}
\end{equation}
where
\[
\begin{split}
 & K^{\mc}_{Gq}(z)
 = \frac{\alpha_s}{2\pi}\, C_F 
   \left\{ \frac{1 + (1-z)^2}{z}\ln\frac{(1-z)^2}{z} + z\right\},
\\& 
 K^{\mc}_{GG}(z) =
 \frac{\alpha_s}{2\pi}\, C_A 
 \Bigg\{
  4\left[\frac{\ln(1-z)}{1-z} \right]_+ 
  + 2\left[\frac{1}{z} - 2 + z(1-z) \right]\ln\frac{(1-z)^2}{z}  
  - 2 \frac{\ln z}{1-z} 
\\&~~~~~~~~~~~~~~~~~~~~~~~~~~~~~~~~~
  - \delta(1-z) \left( \frac{\pi^2}{3} + \frac{341}{72} 
  - \frac{59}{36}\frac{T_f}{C_A}\right)
 \Bigg\},
\\&
K^{\rm MC}_{qq}(z)
= \frac{\alpha_s}{2\pi}\,C_F \,
 \Bigg\{
   4\left[\frac{\ln(1-z)}{1-z}\right]_+ 
   - (1+z) \ln \frac{(1-z)^2}{z} - 2 \frac{\ln z}{1-z} + 1 - z   
   - \delta(1 - z)\left(\frac{\pi^2}{3} + \frac{17}{4}\right)
 \Bigg\},
\\&
K^{\rm MC}_{qG}(z)
= \frac{\alpha_s}{2\pi}\,T_R \,
 \left\{ \left[z^2 + (1-z)^2\right]\ln \frac{(1-z)^2}{z}  + 2z(1-z) \right\}.
 \end{split}
\]
All virtual parts $\sim\delta(1-z)$ are adjusted using the PDFs momentum sum rules.

Alternatively, direct fitting of the LO PDFs 
in the MC scheme directly to DIS data
requires the following NLO coefficient functions
of the DIS process in the MC scheme:
\begin{equation}
\begin{split}
&
C^{\mc}_{2,qq}(z)  =
\frac{\alpha_s}{2\pi}\,C_F \,
\Big\{\Big[ -\frac{1+z^2}{1-z}\ln(1-z) 
   - \frac{3}{2} \frac{1}{1-z} + 3z + 2\Big]_+ + \frac{3}{2}\,\delta(1-z)
\Big\},
\\&
C_{2,qG}^{\mc}(z) =
\frac{\alpha_s}{2\pi}\,T_R \,
\Big\{ -\left[ z^2 + (1 - z)^2 \right] \ln(1-z)  + 6z(1-z)  - 1 \Big\},
\end{split}
\end{equation}
instead of the following ones of the $\msbar$ scheme:
\begin{equation}
\begin{split}
& 
C^{\msbar}_{2,qq}(z) =
\frac{\alpha_s}{2\pi}\,C_F \,
\Big[ \frac{1+z^2}{1-z}\ln\frac{1-z}{z} 
     - \frac{3}{2} \frac{1}{1-z} + 2z + 3\Big]_+ ,
\\&
C^{\msbar}_{2,qG}(z) =
\frac{\alpha_s}{2\pi}\,T_R \,
\Big\{ \left[ z^2 + (1 - z)^2 \right] \ln\frac{1-z}{z}  
    + 8z(1-z) - 1 
\Big\}.
\end{split}
\end{equation}

Let us discuss briefly some important issues
concerning the definition and the use of the MC factorization scheme.
What is the main purpose of the MC factorization scheme?
It is defined such that $\Sigma(z)\delta(k_T)$ terms due to emissions
from initial partons disappear completely from the exclusive NLO corrections.
Such terms exist in the fixed-order NLO calculations independently
of matching with the PS MC%
\footnote{ For instance in ref.~\cite{Catani:1996vz}
   they are termed as ``collinear remnants''.}.
It is vital to remove them in the \krknlo\ scheme, because
without eliminating them it is not possible to include the NLO corrections
using simple multiplicative MC weights on top of the PS MC distributions.

How to determine elements of the transition matrix  $K^{\mc}_{ab}$?
They can be be obtained from inspecting NLO corrections to two processes
with initial-state quarks and gluons in the LO hard process:
$Z/\gamma$  and Higgs-boson production, 
as was done in ref.~\cite{Jadach:2016acv}.

The immediate question, of course, is
whether the same set of PDFs in the MC factorization scheme is able
to eliminate unwanted $\sim\delta(k_T)$ terms for other processes?
This is a question about {\em universality} of the MC factorization scheme.
It was shown already in ref.~\cite{Jadach:2011cr},
that this is true for the DIS process.
Preliminary analysis shows that
for any other process with any number of colored particles it may work,
provided PS MC distributions
(or the corresponding soft-collinear counter terms in the NLO calculations)
obey certain extra conditions.
The corresponding discussion will be reported/published in a separate study.

\subsection{On the quality of LO PS MC}
In the implementation of other methods 
of combining (matching) NLO-corrected hard process
with LO PS,
the MC generation of the distribution of the hardest parton
responsible for the bulk of the LO+NLO corrections
(\powheg~\cite{Nason:2004rx}) or missing NLO (\mcatnlo~\cite{Frixione:2002ik})
is taken outside LO PS MC and is done using separate software tools,
before LO PS MC is invoked to add more partons.
In \krknlo\ the order is inverse, PS MC is invoked first to generate
the hardest parton and other partons together and then the NLO corrections
are introduced in the next step by means of reweighting the generated MC events.

Assuming for simplicity the $k_T$-ordering in PS MC, the distribution
of the hardest parton generated by LO PS MC has to be controlled
quite strictly, that is analytically%
\footnote{This is true for other methods as well.}.
It has also to reproduce properly soft and collinear limits
of the NLO distributions.
For simple processes
with two initial-state hadrons and any color-neutral object in the final state, 
like DY or Higgs-boson production,
the above requirements are easily fulfilled by any modern PS MCs.
In ref.~\cite{Jadach:2015mza} which implements the \krknlo\ method 
for the DY process
it was checked very carefully that the LO distributions of
\sherpaver\ and \herwigver\ are identical with the
soft-collinear counter terms of the Catani--Seymour 
subtraction scheme~\cite{Catani:1996vz}.

Another related critical point is the use of the backward
evolution (BEV) algorithm in LO PS MC.
It is easy to prove that the LO backward and forward evolutions
are the same up to LO.
However, the control of the NLO terms in the BEV algorithm is
less trivial, especially for the terms due to kinematic limits.
This problem was analyzed in a detail in ref.~\cite{Jadach:2015mza},
where it was shown that such a strict NLO-level control of the
distributions generated by BEV algorithm is possible.
In particular, the role of the veto algorithm in BEV in assuring
the full coverage of the phase space up to $k_T=\sqrt{s}/2$ was clarified.

In the case of more than three colored partons in the LO hard process,
the requirement on PS MC that it reproduces exactly the QCD soft-collinear limit
is difficult to meet and typical LO PS MCs implement some approximate 
distributions, which have to be corrected once the NLO corrections
are implemented using the \krknlo\ method
(provided the full coverage of the soft gluon phase space is assured).

\begin{figure}[ht]
\centering
\includegraphics[height=0.27\textheight,width=0.48\textwidth]{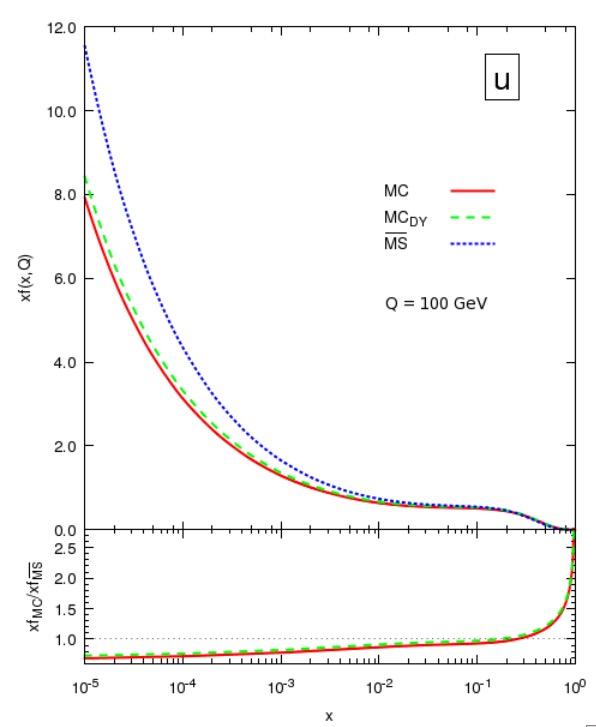}
\includegraphics[height=0.27\textheight,width=0.48\textwidth]{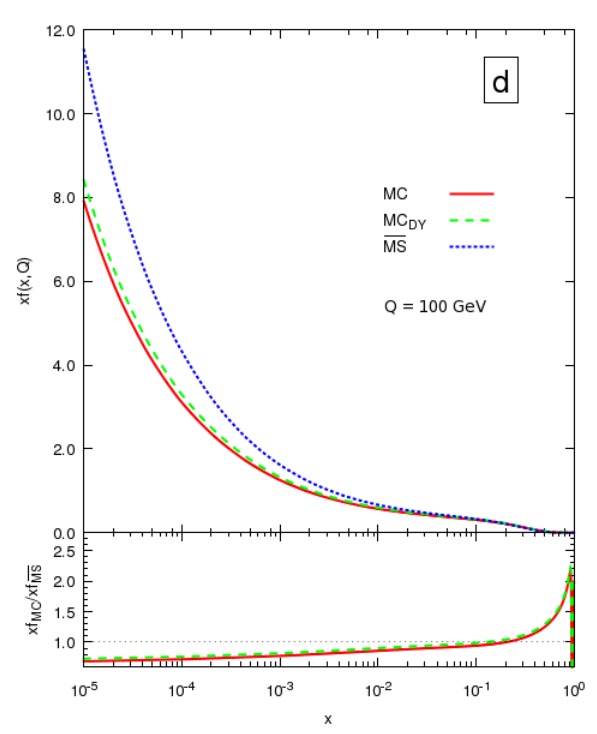}

\includegraphics[height=0.27\textheight,width=0.48\textwidth]{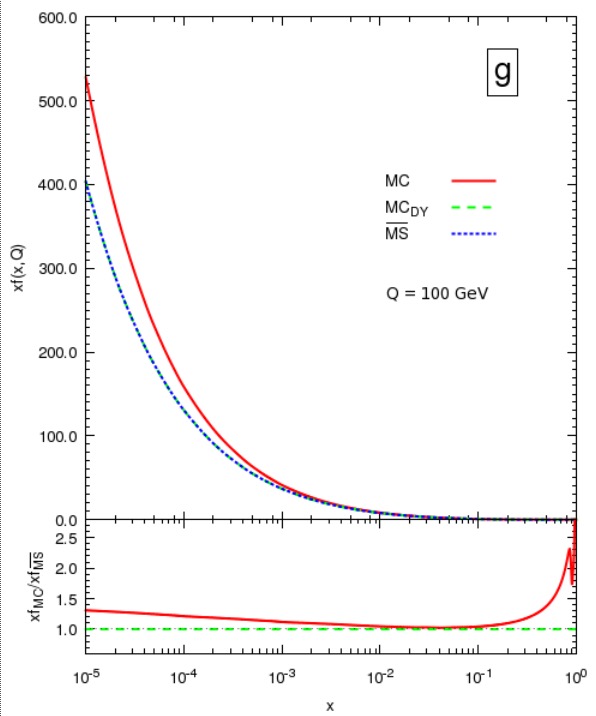}
\includegraphics[height=0.27\textheight,width=0.48\textwidth]{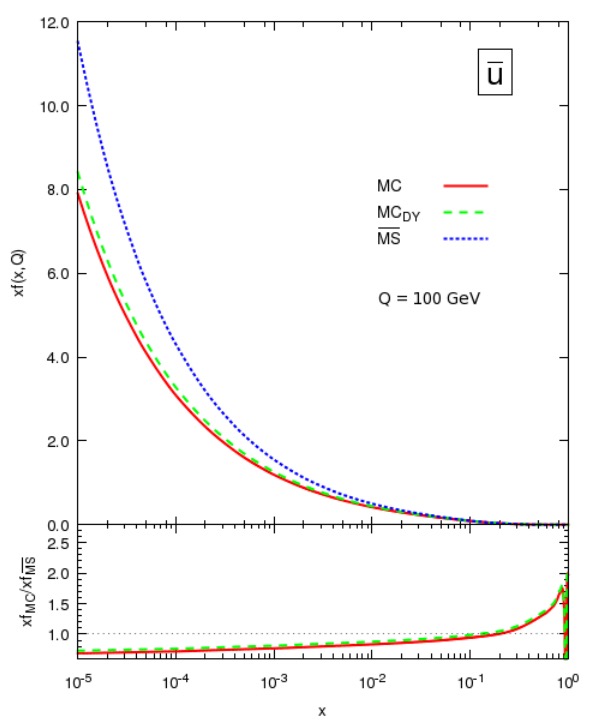}

\caption{\sf
   Examples of PDFs in the MC factorization scheme.
   }
\label{fig:PDFs}
\end{figure}

\subsection{Numerical examples of PDFs in the MC factorization scheme}
In Fig.~\ref{fig:PDFs} we show a few examples of PDFs transformed 
from the $\msbar$ to the MC factorization scheme.
They will be used in the following section
for  the Higgs-production process.

As seen in Fig.~\ref{fig:PDFs}, the 
change of PDFs in the MC scheme with respect to the $\msbar$ is noticeable.
Lines labeled with MC represent the complete MC scheme
as defined in eq.~(\ref{eq:msbar2mc}).
Lines marked with MC$_{DY}$ use simplified $K$-matrix 
of ref.~\cite{Jadach:2015mza}.
This simplification was acceptable for the DY process
modulo ${\cal O}(\alpha_s^2)$ terms.
Generally, for low $x$ the gluon PDF is smaller for the $\msbar$ than for the MC scheme,
while for quarks it is the other way around.

%

\section{Higgs-production process}
The application of the \krknlo\ method for the DY process was covered in a great
detail in ref.~\cite{Jadach:2015mza},
where numerical results for the total cross section, distributions of
rapidity and transverse momentum
were shown for the \krknlo\ method,
in comparison with fixed-order NLO of \mcfm~\cite{MCFM},
NNLO of ref.~\cite{Catani:2009sm} and also with the results
of the \powheg~\cite{Nason:2004rx} 
and \mcatnlo~\cite{Frixione:2002ik} implementations.

Similar extensive panorama of numerical results for the Higgs-production
process using the \krknlo\ method 
will be published soon~\cite{Jadach:2016tobepub}.
Here we are going to show preliminary results for an integrated cross section.
In the table below we include the results
for the total cross section of the Higgs production in pp collisions 
at $\sqrt{s}=8$~TeV from the \krknlo\ method
compared with that of \mcatnlo~\cite{Frixione:2002ik}:
\begin{center}
\begin{tabular}{l|l}
 & $\sigma_{tot} [pb]$    \\
\hline
\mcatnlo\ & $18.72 \pm 0.04$     \\
\krknlo\  & $19.38 \pm 0.04$     \\
\end{tabular}
\end{center}
The difference of $\sim 3.5\%$ can be attributed 
to different treatment in the two matching methods of effects that are
beyond the strict NLO approximation.
 

In the preliminary  results,
to be shown in ref.~\cite{Jadach:2016tobepub},
the rapidity distributions
from the \krknlo\ method implemented on top of the most recent
version of \herwigsiedem~\cite{Bellm:2015jjp},
differs by $\leq 20\%$ from that of
\mcatnlo~\cite{Frixione:2002ik} methods,
well within uncertainty typical for NLO.
A similar difference, typical for NLO, was found
in transverse momentum distribution for $p_T\leq 100\,$GeV.
For higher transverse momenta
$\sim 200\%$ differences with \mcatnlo\ are found, 
well known from comparisons between
\powheg\ and \mcatnlo\ for this process, see ref.~\cite{Alioli:2008tz}.

\section{Summary and outlook}
The \krknlo\ method of introducing the NLO QCD corrections
in the LO parton shower Monte Carlo is a simple scenario for
the NLO-corrected PS MC -- an interesting alternative to 
the more complicated \mcatnlo\ or \powheg\ approaches.
Potential gains from the new QCD matching method are:
(i) reducing h.o. QCD uncertainties,
(ii) possibly easier implementation of NNLO corrections.
An interesting application would be a high-quality QCD+EW+QED
Monte Carlo event generator with a hard process like single or multiple 
$W/Z/H$ heavy boson production at high luminosity LHC.
A longer-term possibility is to extend the \krknlo\ technique
to the N+NLO calculation,
with the NLO ladder (PS MC) and NNLO-corrected hard process.


\end{document}